\documentclass[letterpaper,twocolumn,10pt]{article}
\usepackage{usenix2019,epsfig,endnotes}
\usepackage[dvipsnames]{xcolor}

\usepackage{hyperref}
\usepackage{graphicx}
\usepackage{comment}
\usepackage{authblk}
\graphicspath{ {./images/} }

\title{\LARGE \bf
Nature of System Calls in CPU-centric Computing Paradigm
}

\author[1]{Viacheslav Dubeyko}
\author[2]{Om Rameshwar Gatla}
\author[2]{Mai Zheng}
\affil[1]{Western Digital Research}
\affil[2]{Iowa State University}

\begin{document}

\maketitle
\pagestyle{plain}

%%%%%%%%%%%%%%%%%%%%%%%%%%%%%%%%%%%%%%%%%%%%%%%%%%%%%%%%%%%%%%%%%%%%%%%%%%%%%%%%
\begin{abstract}

Modern operating systems are typically POSIX-compliant 
with major system calls  specified decades ago.
%The system calls are the fundamental layer of interaction between user-space applications and the OS kernel and its implementation of fundamental abstractions and primitives used in modern computing.
The next generation of non-volatile memory (NVM) technologies
%NVM/SCM memory 
%raises critical questions about the efficiency of modern OS architecture. 
raise concerns about the efficiency of the traditional POSIX-based systems. % for the new memories. 
As one step toward building high performance NVM systems,
we explore the potential dependencies between system call performance and major hardware components (e.g., CPU, memory, storage)
under typical user cases (e.g., software compilation, installation, web browser, office suite) in this paper. %-space applications
%we attempt to understand the interaction dynamics between user space and kernel space
%for several typical user-space applications (software compilation, installation, diff viewer, Firefox, OpenOffice).
We build histograms for the most frequent and time-consuming system calls with the goal to understand the nature of distribution on different platforms. 
%It was discovered unexpected and stable dependence of histograms from the CPU architecture. On the other hand, the type of persistent storage does not play important role in the discovered dependence.
We find that there is a strong dependency between the system call performance and the CPU architecture.
On the other hand, 
the type of persistent storage plays a less important role in affecting
the performance. %does not play important role. % in the discovered dependence.

\end{abstract}

%%%%%%%%%%%%%%%%%%%%%%%%%%%%%%%%%%%%%%%%%%%%%%%%%%%%%%%%%%%%%%%%%%%%%%%%%%%%%%%%


%%%%%%%%%%%%%%%%%%%%%%%%%%%%%%%%%%%%%%%%%%%%%%%%%%%%%%%%%%%%%%%%%%%%%%%%%%%%%%%%
\section{INTRODUCTION}

%The next generation of NVM/SCM memory would be able to open the new horizons for future computer technologies. However, currently, the NVM/SCM memory represents the big challenge but not the hope to resolve the computer science's problems. Modern Operating Systems are typically POSIX-compliant. POSIX defines the application programming interface (API), along with command line shells and utility interfaces. The libc and libstdc++ are the final boundaries for any user-space applications. These libraries implement a set of primitives that interact with kernel-space via system calls in order to delegate some low-level tasks to the kernel.
The next generation of non-volatile memory (NVM) or storage class memory (SCM) technologies %~\cite{ReRAM,STT-RAM,PCM-IBM}
are expected
to provide durability similar to the flash memory
with latencies comparable to DRAM. These unique
characteristics bring new concerns on the efficiency of modern POSIX-compliant operating systems, since the major interface was  specified decades ago.

%In this work, we attempt to understand the interaction dynamics between user space and kernel space for several typical user-space applications (software compilation, installation, diff viewer, Firefox, OpenOffice). 
%We studied the strace output for several platforms (AMD E-450, i7-3630QM, XEON E5-2620 v2) with the Ubuntu Linux distribution (3.13.0-24, 3.13.0-95, 4.2.0-42 kernel versions) on different storage devices HDDs (5200 RPM SATA 3.0, 15015 RPM SAS), SSD (SATA 3.1, PCIe) and RAMdisk while using different file systems (ext2, ext4, XFS, tmpfs). We built histograms for the most frequent and time-consuming system calls with the goal to understand the nature of distribution for different platforms. 
%We discover unexpected and stable dependence of histograms from the CPU architecture. However, the type of persistent storage plays a less important role in the system call performance. %the discovered dependence. 

In this paper, we explore the potential dependency 
between POSIX system call performance and major hardware components.
We monitor the 
 system calls under 
five representative use cases (e.g., software compilation, installation, diff viewer, Firefox, OpenOffice) through the \texttt{strace} utility~\cite{c121},
calculate their distributions,
and identify the ones that are used most frequently and 
are most time-consuming.
Moreover,  we built histograms for the major system calls 
on six platforms
with a wide range of CPUs (e.g., AMD E-450, i7-3630QM, XEON E5-2620 v2),  storage devices (e.g.,  15015 RPM SAS HDD, SSD SATA 3.1 SSD, PCIe SSD, RAMdisk),
OS kernels (e.g., 3.13.0-24, 3.13.0-95, 4.2.0-42),
and  file systems (e.g., ext4, XFS, tmpfs).
Through  cross-platform analysis,
we 
find that there is a strong dependency between the system call  performance  and  the  CPU  architecture.   
On  the  other hand,  the type of persistent storage plays a less important role in the system call performance. %the discovered dependence. 
We hope this study could help understand the performance bottleneck of existing
system call implementations, and thus facilitate building high performance NVM systems.

The rest of this paper is organized as follows. Section II presents the background and motivation of the paper. Section III surveys the related works. Section IV explains the methodology (experimental setup, use-cases, research tool) that was used in this work. Section V contains use-cases analysis on the basis of total frequency (number of calls) and total time consumption of different system calls. Section VI introduces the system calls analysis on the basis of prepared histograms. Section VII includes final discussion of the system calls analysis. Section VIII offers conclusions.

\section{Background \& Motivation}

\textbf{Next generation of NVM memory} is able to provide a persistent byte-addressable space for storing, accessing and modification of data. The features of the new persistent memory are very attractive and to promise the potential resolving of many critical problems in the current data processing paradigm.

\noindent
\textbf{CPU vs. NVM memory}. Existing hardware and software stack has huge amount of issues that prevent from efficient using of NVM/SCM memory in current computing systems. First of all, modern CPUs are not ready to access the byte-addressable space of NVM memory. Another critical point is inability of modern CPUs to manage the persistent nature of NVM memory. The persistency of NVM memory is the issue but not the advantage for the modern CPUs. Current state of the art of hardware stack includes several levels of memory hierarchy (CPU's registers, L1/L2/L3 caches, DRAM, persistent storage). This architecture is the compromise that took place because of block-based interface and huge latency of regular persistent storage devices (HDD, SSD). However, NVM memory is able to suggest the low access latency, byte-addressable nature and persistence. If anybody considers to use the NVM memory in the persistent storage device then all advantages of NVM memory (byte-addressable nature and low latency) will be neutralize by the latency of storage device's controller and by the overhead of block-based interface of OS's software stack.

\noindent
\textbf{NVM memory like main memory}. The idea to use the NVM memory instead of DRAM like the main memory reveals the many drawbacks too. Many NVM memory technologies have not very high endurance. It means that the direct CPU access is able to kill the NVM chip very quickly. Usually, the main memory is volatile and as CPU as OSes is unable to use the persistent nature of NVM memory like substitution of DRAM. Moreover, the persistence of NVM memory can be a source of various security issues for the case of using like the main memory. There are significant number of research efforts with the goal to suggest the new hardware-based approaches or OS's subsystems modification for efficient using of NVM memory. However, all these efforts have one critical downside. The efficiency of every approach is estimated by means of special benchmarks that are unable to simulate the real life. It means that any benchmark is unable to show the real efficiency of computing system in the real environment (for example, for the case of typical environment of regular end-user).

\noindent
\textbf{System calls analysis}. Usually, OS is split on user-space and kernel-space. The user-space is a world of applications that can request some services from hardware devices by means of system calls are provided by kernel-space. Generally speaking, any hardware resources are managed by kernel space and user-space application is able to request any services by means of system calls. Finally, the NVM memory is hardware resource that should be managed by kernel-space and is available for user-space application through the system calls. It means that system calls analysis is the fundamental way to achieve the understanding how efficiently the whole computing system can operate. It is possible to build the testing space by means of variation of hardware resources (CPUs, DRAM capacity, storage devices) and use-cases. This testing space is able to reveal the key peculiarities and tendencies of the computing system at whole. The analysis of system calls' frequency and time consumption is able to show the important peculiarities of modern OSes. These results can be the basis for understanding the key bottlenecks of modern OSes that can be used for elaboration of vision(s) how the next generation of NVM memory can really be used efficiently.

\section{Related works}

\textbf{System support for NVM}. Great efforts have been put towards providing software and/or hardware support for NVM \cite{c100,c101,c102,c104,c105,c106}. For example, Mnemosyne \cite{c104} and NVML \cite{c100} provide transaction libraries for durable transactions and atomic updates on NVM. PMFS \cite{c102}, NOVA \cite{c105} and NOVA-Fortis \cite{c106} enable accessing NVM through the file system interface. Our work is complementary to these efforts in that we focus on identifying the fundamental bottleneck of existing system call interfaces and implementations, which is critical for re-designing systems for NVM.

\noindent
\textbf{Performance Analysis of NVM-based systems}. Many researchers have analyzed the performance of NVM-based systems and/or applications \cite{c100,c101,c102,c103,c104,c105,c106,c107}.  For example, WHISPER \cite{c103} quantitatively evaluates 10 NVM-aware benchmarks using three access interfaces (i.e., native, library, and file system) and identify important characteristics of NVM applications (e.g., software transactions are often implemented with 5 to 50 ordering points).  Different from these existing efforts which mostly uses one type of hardware (e.g., DRAM or hardware emulator) and a limited number of workloads, we cover a wide spectrum of hardware platforms, use cases, and system calls.

%\textbf{Performance Analysis of Non-NVM Systems}. XXX

\noindent
\textbf{System Calls Optimization}. The inefficiency of system calls was discussed in many papers \cite{c108,c109,c110,c111,c112,c113,c114,c115}. Triplett et al \cite{c110} proposed rethinking the division between user-space and kernel space to eliminate the overhead of system calls. Rather than dedicating only a single core to a process, they suggested to dedicate two: one to run the user-mode process, and one to perform system calls and the associated kernel-mode computation. Soares et al \cite{c112} proposed exception-less system calls. They showed that synchronous system calls negatively affect performance in a significant way, primarily because of pipeline flushing and pollution of key processor structures (e.g., TLB, data and instruction caches, etc.). In their implementation, system calls are issued by writing kernel requests to a reserved syscall page, using normal memory store operations. The actual execution of system calls is performed asynchronously by special in-kernel syscall threads, which post the results of system calls to the syscall page after their completion. Rajagopalan et al \cite{c114} suggested a profile-directed approach to optimizing a program’s system call behavior (system call clustering). In this approach, profiles are used to identify groups of systems calls that can be replaced by a single call, thereby reducing the number of kernel boundary crossings. They showed an average 25\% improvement in frame rate, 20\% reduction in execution time, and 15\% reduction in the number of cycles for the mpeg play video software decoder.

\noindent
\textbf{System Analysis via System Calls Tracing}. Tracing the system calls is very useful technique, especially, for the case of debugging an application. The researchers use this technique for analysis of system efficiency at whole \cite{c116,c118,c119,c120}. Park et al \cite{c116} evaluated battery performance on the Android platform by tracing system calls. Kodirov et al \cite{c119} used system calls tracing for investigation of the text editing program gedit and the libraries it relies upon. They identified in application the recurring patterns that can be simplified if better FS support were available.

\section{Methodology}
As one step towards building high-performance NVM systems,
we explore one important research question:
%Is the latency of popular system calls affected by 
%the type of persistent storage? 
Is the execution time of popular system calls mainly affected by 
the type of persistent storage? 
If so, how much? If not, what is the major factor affecting the execution time?

To answer the question, we monitor the 
 system calls under 
five representative use cases through the \texttt{strace} utility~\cite{c121},
calculate the distribution of system calls in terms of frequency,
and identify the ones that are used most often.
Moreover, we record the execution time of each system call invocation, 
and compare them across six platforms with different hardware components.

\begin{comment}
To answer the question, we monitor 
five representative use cases through the \texttt{strace} utility~\cite{c121}, and record the frequency of each system call involved. 
We calculate the distribution of system calls in terms of frequency
and identify the ones that are used most often.
Moreover, we record the latency of each system call invocation, 
and compare them on six platforms with different persistent storage 
components.
\end{comment}

\begin{comment}
As one step towards building high-performance NVM systems, 
we analyze the existing system call implementations under
representative workloads on different platforms in this study.
We explore four research questions: (1) what system calls are involved in typical use cases? (2) what are the distributions of the system calls involved? (3) what are the most time-consuming system calls? (4) 
Does the performance of system calls vary across different hardware?

Specifically, given an application and a system, 
we use the \texttt{strace} utility~\cite{c121} to intercept 
and record the system calls invoked by the application, measure the frequency and latency of each system call, and identify the variability across platforms.
To make the study comprehensive and representative, 
we use a variety of platforms and use cases as described in the following two subsections.
\end{comment}

\subsection{Experimental Platforms}

%The most critical artifacts of the computing paradigm are: (1) CPU, (2) DRAM, (3) storage device, (4) file system.

%\begin{figure}[h]
%\centering
 
\begin{table*}[tb]
    \begin{center}
    \begin{tabular}{c | c| c |c | c| c |c | c| c}
      \hline
  %    \multicolumn{3}{c|}{ --} & Platform\#1 & Platform\#2  &Platform\#3   & Platform\#4 & Platform\#5 & Platform\#6 \\   
  \multicolumn{3}{c|}{ --} & Platform 1 & Platform 2  &Platform 3   & Platform 4 & Platform 5 & Platform 6 \\ 
      \hline
      \hline
    % \texttt{cp+tar+rm}     & copy, compress/decompress, \& delete files & age Lustre \\ 

       &     &type & AMD E-450 &AMD E-450 & Intel i7  & Intel Xeon  &Intel Xeon & Intel Xeon \\ 
       &     &      & 1,650 MHz& 1,650 MHz &  3630QM    & E5-2620 v2  &E5-2620 v2 &E5-2620 v2  \\ 
       &     &      &         &          & 2.40 GHz   & 2.10 GHz    &2.10 GHz  &2.10 GHz   \\ 

             \cline{3-9} 
       & CPU &core & 2& 2& 8 &24  &24 & 24 \\ 
             \cline{3-9} 
       &     & cache     &L1 32 KB  &L1 32 KB &L1 32 KB  &L1 32 KB  &L1 32 KB &L1 32 KB  \\ 
       &     &  &L2 512 KB &L2 512 KB &L2 256 KB  &L2 256 KB  &L2 256 KB  &L2 256 KB   \\ 
       &     &      & & &L3 6 MB   &L3 15 MB   &L3 15 MB  &L3 15 MB   \\
     \cline{2-9} 
       &       &type &SODIMM &SODIMM &SODIMM  &DIMM  &DIMM &DIMM  \\
    HW   & Memory&     &DDR3 &DDR3 &DDR3  &DDR3  &DDR3 &DDR3  \\
       &       &     &1,333 MHz   &1,333 MHz    &1,600 MHz  &1,866 MHz  &1,866 MHz  &1,866 MHz   \\
       \cline{3-9} 
       &      &size & 8 GB & 8 GB &24 GB  & 16 GB &16 GB & 16 GB \\
     \cline{2-9} 
       &         &type &HDD &SSD &SSD  &HDD  &SSD &DRAM  \\
                 \cline{3-9} 
       &Persistent&RPM &5,200 & --& -- & 15,015 & -- & -- \\
                  \cline{3-9} 
       &Storage   &size &2 TB &500 GB &480 GB  &73.4 GB  &128 GB &16 GB  \\
                  \cline{3-9} 
       &       &protocol & SATA 3.0 &SATA 3.1 &SATA 3.1  &SAS  &PCIe &--  \\
       &       &        & 6 Gb/s  &6 Gb/s &6 Gb/s  &6 Gb/s &6 Gb/s &  \\
\hline
\hline
 %  & \multicolumn{2}{c|}{Kernel} & Linux & Linux& Linux &Linux  &Linux &Linux  \\
OS &  \multicolumn{2}{c|}{Linux kernel version} & 3.13.0-24 & 3.13.0-24  &3.13.0-95  & 3.13.0-24  &3.13.0-24  &4.2.0-42  \\
\cline{2-9} 

   &   \multicolumn{2}{c|}{File system} & Ext2 &Ext4 &XFS & Ext2 & XFS &tmpfs  \\
  
     % \texttt{WikiW-sync}     &   write new  files  synchronously, read back  \&  verify MD5 & analyze post-LFSCK behavior \\  
      \hline
    \end{tabular}
    \end{center}
    \vspace{-0.15in}
    \caption{
  {\bf Six platforms with different hardware (HW) and operating systems (OS).} }
\label{tab:platforms}
\end{table*}

%Fig. \ref{fig:fig001}
%Table~\ref{tab:platforms} contains description of all hardware platforms used in the experiments. We selected several CPUs with: (1) different architectures (AMD E-450, XEON E5-2620, i7-3630QM), (2) one (AMD, i7) and two (XEON) physical sockets; (3) various core numbers (2 -– 24 cores); (4) various L1/L2/L3 cache sizes. We varied the DRAM size from 8 GB up to 24 GB. We used several HDDs (5200 RPM SATA 3.0, 15000 RPM SAS) and SSDs (SATA 3.1, PCIe). The RAMdisk was chosen as a special case that approximates an “ideal” environment in order to estimate the efficiency of POSIX-based OS in an environment with fast persistent memory. The ext2, ext4 and xfs file system were selected because of their support for Direct Access (DAX). The tmpfs file system was used for the case of RAMdisk based platform.

We use six platforms with a wide range of hardware and software. 
As shown in Table~\ref{tab:platforms},
we select several CPUs with: (1) different architectures (AMD E-450, Intel Xeon E5-2620, Intel i7-3630QM); (2) various core numbers (2 -– 24); (3) various L1/L2/L3 cache sizes. Similarly, we vary the DRAM size from 8 GB up to 24 GB. 
In terms of persistent storage, 
we use several HDDs (5200 RPM SATA 3.0, 15000 RPM SAS) and SSDs (SATA 3.1, PCIe). 
Since persistent memory is immature and not publicly available, 
we use RAMdisk 
to approximate an “ideal” persistent memory environment. %to evaluate the efficiency of POSIX-based OS in an environment with fast persistent memory.
In terms of operating systems, we use the Ubuntu distribution with 
kernel versions 3.13.0-24 and 4.2.0-42.
 The Ext2, Ext4 and XFS file system are selected because of their support for Direct Access (DAX). 
 The tmpfs file system is used for the RAMdisk based platform.
 
\subsection{Use Cases}

%We needed to choose such use-cases that exhibit: (1) critical bottlenecks of the whole system; (2) potential directions for performance improvements.

%\begin{figure}[h]
%\centering
% \includegraphics[width=0.90\columnwidth,keepaspectratio]{table2-use-cases}
%\caption{Use cases.}
%\label{fig:fig002}
%\end{figure}

\begin{table}[tb]
    \begin{center}
    \begin{tabular}{c | c}
      \hline
      Use Cases   &  Description \\   
      \hline
      \hline
    % \texttt{cp+tar+rm}     & copy, compress/decompress, \& delete files & age Lustre \\ 
    \texttt{Compilation}   &  compile F2FS utilities via    \\ 
    &  \texttt{autoreconf}, \texttt{configure}, \& \texttt{make}    \\ 
    \hline
  \texttt{Installation}     &  install  applications via   \\ 
           &   Ubuntu Software Center    \\ 
\hline
        \texttt{FireFox}   & a web browser   \\ 
     \hline
       \texttt{OpenOffice Calc}      &  a spreadsheet application  \\ 
             \hline

             \texttt{Meld}   & a visual diff and merge tool  \\  
         \hline

    \end{tabular}
    \end{center}
    \vspace{-0.15in}
    \caption{
    {\bf Five representative use cases.} 
   % {\it This table shows the five use cases used in our study. } 
    }
\label{tab:usecases}
\end{table}

%As a result, we analyzed the following use-cases %(Fig. \ref{fig:fig002}): 
%Table~\ref{tab:usecases} shows the five typical use cases for experiments:
%(1) software installation, (2) diff viewer, (3) internet browser, (4) OpenOffice, (5) compilation. These use-cases were selected as typical cases of end users usage scenarios in a desktop system. Software installation was expected to be dominated by write operations, the diff viewer use case by metadata operations, the internet browser by operations on lots of small files, OpenOffice to be dominated by read operations, and compilation a mixed case of read/write/metadata operations. The test workflow included: (1) start OS, (2) synchronize data on disk with memory, (3) drop all caches, (4) run use-case with strace logging, and, (5) optional window termination. The goal of the research was: (1) evaluate application profile – what system calls are dominant in that specific use-case; (2) evaluate the most time-consuming system calls in the use-case; (3) detect changes in time consumption on different hardware platforms for different use-cases.

As shown in Table~\ref{tab:usecases}, 
we select five representative use cases to drive the target systems: 
\texttt{Compilation} configures and compiles F2FS utilities~\cite{c122} through a set of common compiler tools including  \texttt{autoreconf}, \texttt{configure}, and \texttt{make};
\texttt{Installation}  installs multiple applications through the Ubuntu Software Center;
 the remaining three use cases (i.e., \texttt{FireFox}, \texttt{OpenOffice Calc}, and \texttt{Meld}) run popular applications for 
 web browsing, spreadsheet calculation, and diff viewing respectively,
 all of which involve user interactions (e.g., browsing 
 across multiple websites and clicking links).
 Together, the five cases cover typical usage scenarios of end users in a desktop environment.
 
\begin{comment}
To expose the performance bottlenecks and to identify the directions for improvement, 
we select five representative use cases to drive the target systems. 
As shown in Table~\ref{tab:usecases}, the use cases include
three popular applications (
\texttt{FireFox}, \texttt{OpenOffice Calc}, and \texttt{Meld}) and
two synthetic workloads (\texttt{App-Compile}
and \texttt{App-Install}).
The three applications involve sophisticated user interaction (e.g., browsing 
across multiple websites and clicking links in \texttt{FireFox}).
\texttt{App-Compile}  configures and compiles F2FS utilities~\cite{f2fs-utilities} through a common compiler tool chain (i.e., autoreconf, configure, make).
\texttt{App-Install} installs multiple applications through the Ubuntu Software Center.  
Together, the five cases cover typical usage scenarios of end users in a desktop environment.
\end{comment}

Moreover, the five use cases are expected to cover different I/O patterns. 
For example, 
\texttt{Compilation} 
involves a mix of read/write operations,
\texttt{Installation} is dominated by write operations, 
and \texttt{Meld} involves many file system metadata operations for 
traversing directories and merging files.

In addition, to minimize the impact of buffering in the target systems, 
we explicitly flush buffers and drop the dirty pages before running each use case 
(i.e., using \texttt{sync} and \texttt{echo 3> /proc/sys/vm/drop\_caches}).

\section{Distribution of System Calls}

%The frequency and time consumption factors revealed that
%autoreconf, configure and make use-cases 
%(Fig. \ref{fig:fig003}) distribute the execution activity among:
To identify the most popular and the most
time-consuming system calls, we calculate the system call distribution under different use cases
based on the usage frequency.

\begin{figure}[h]
\centering

\includegraphics[width=0.7\columnwidth,keepaspectratio]{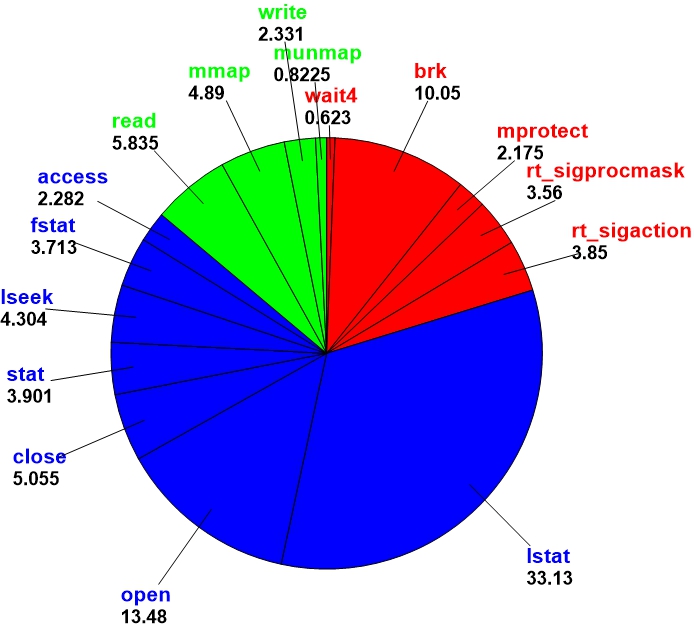}
\caption{%System calls frequency distribution for make use-case. 
System call distribution for the \texttt{make}
utility in the \texttt{Compilation} use case. OS-related system calls are marked by red color. Metadata-related system calls are marked by blue color. User data related system calls are marked by green color.}
\label{fig:fig003}
\end{figure}

Figure~\ref{fig:fig003} shows the distribution of system calls 
for the \texttt{make} utility in the \texttt{Compilation} use case.
We can see that \texttt{make}
distributes the execution activity among 
a variety of system calls including: (1) OS related system calls (mprotect, brk, rt\_sigaction, rt\_sigprocmask, rt\_sigreturn, pipe, dup2, wait4), (2) metadata related system calls (stat, fstat, lstat, open, close, access, lseek), (3) user data related system calls (mmap, munmap, read, write).

\begin{figure}[h]
\centering
 
 \includegraphics[width=0.70\columnwidth,keepaspectratio]{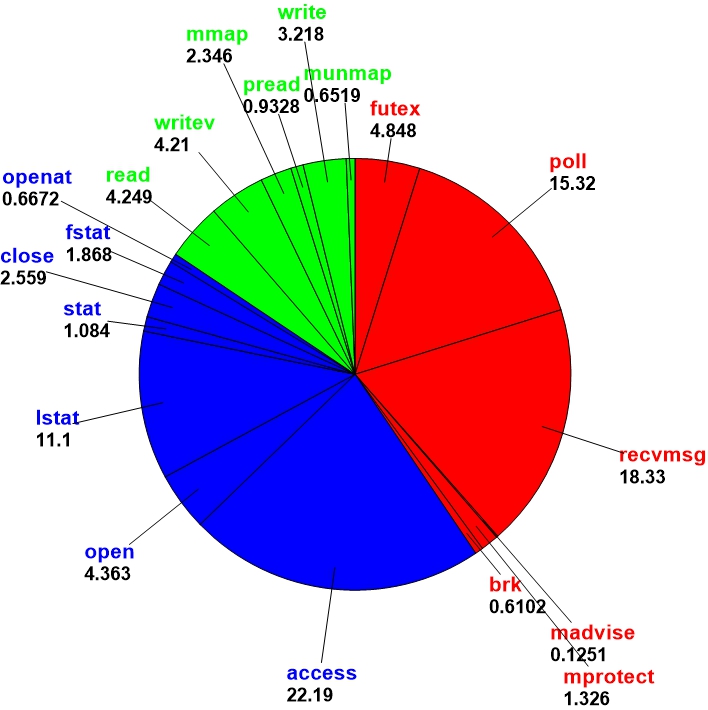}
\caption{System calls frequency distribution for OpenOffice Calc use-case. OS-related system calls are marked by red color. Metadata-related system calls are marked by blue color. User data related system calls are marked by green color.}
\label{fig:fig004}
\end{figure}
\begin{figure}[h]
\centering
 
 \includegraphics[width=0.70\columnwidth,keepaspectratio]{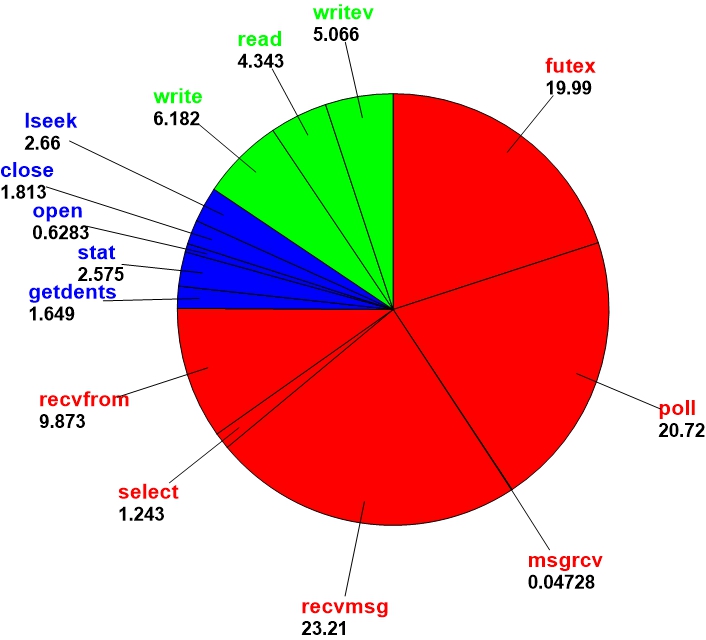}
\caption{System calls frequency distribution for Firefox use-case. OS-related system calls are marked by red color. Metadata-related system calls are marked by blue color. User data related system calls are marked by green color.}
\label{fig:fig005}
\end{figure}
\begin{figure}[h]
\centering
 
 \includegraphics[width=0.70\columnwidth,keepaspectratio]{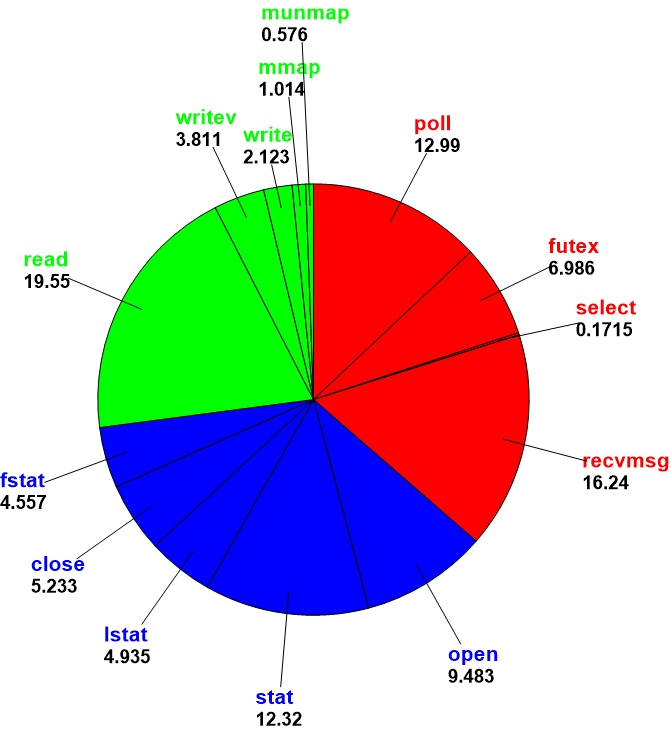}
\caption{System calls frequency distribution for Meld use-case. OS-related system calls are marked by red color. Metadata-related system calls are marked by blue color. User data related system calls are marked by green color.}
\label{fig:fig006}
\end{figure}
\begin{figure}[h]
\centering
 
 \includegraphics[width=0.45\columnwidth,keepaspectratio]{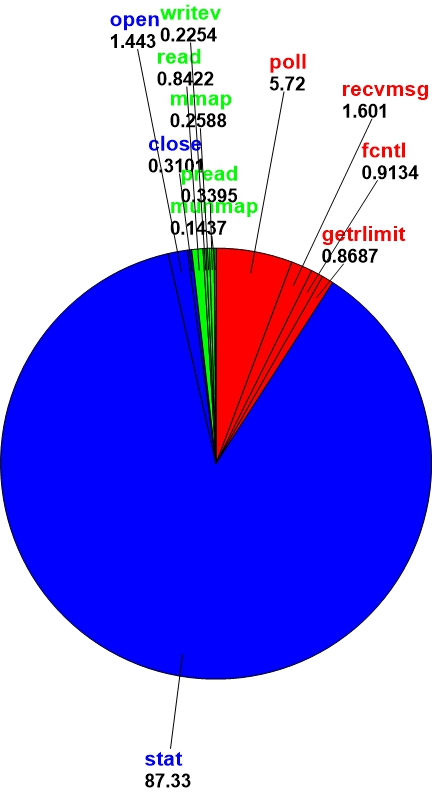}
\caption{System calls frequency distribution for Software Installation use-case. OS-related system calls are marked by red color. Metadata-related system calls are marked by blue color. User data related system calls are marked by green color.}
\label{fig:fig007}
\end{figure}

%\begin{figure}[h]
%\centering
% \includegraphics[width=0.70\columnwidth,keepaspectratio]{table3}
%\caption{Most important system calls.}
%\label{fig:fig008}
%\end{figure}

\begin{table}[tb]
    \begin{center}
    \begin{tabular}{c | c}
      \hline
      Group   &  System Calls \\   
      \hline
      \hline
    % \texttt{cp+tar+rm}     & copy, compress/decompress, \& delete files & age Lustre \\ 
      \texttt{Memory }   & mprotect(), brk(), mmap(),  \\ 
      \texttt{Operations }   & munmap(), madvise()  \\  
\hline
        \texttt{Signal}   & rt\_sigaction(), rt\_sigreturn() \\ 
       \texttt{Processing}   & rt\_sigprocmask()  \\ 

    \hline
       \texttt{Interprocess}      &  pipe(), recvmsg(),  \\   
         \texttt{Communication}   &  recvfrom()  \\ 
    \hline
         \texttt{File}     &  dup2(), stat(), fstat(), lstat(), open()    \\ 
         \texttt{Operations}     &   close(), access(), lseek(), fcntl()    \\ 
    \hline
         \texttt{User Data}     &  read()   \\ 
         \texttt{Operations}     &  write()  \\ 
   \hline
         \texttt{Locking}     &  futex, poll, wait4   \\ 
         \texttt{Operations}     &     \\ 
      \hline
    \end{tabular}
    \end{center}
    \vspace{-0.15in}
    \caption{
    {\bf Summary of the most frequent and time-consuming system calls.} 
   % {\it This table shows the five use cases used in our study. } 
    }
\label{tab:syscall}
\end{table}

Similarly, we measure the distributions of system calls  for other use cases. \texttt{OpenOffice Calc}  (Fig. \ref{fig:fig004}) has some peculiarities in the execution activity distribution but it uses: (1) OS related system calls (mprotect, brk, madvise, recvmsg, futex, poll), (2) metadata related system calls (stat, fstat, lstat, open, close, access), (3) user data system calls (mmap, munmap, read, write). 
\texttt{Meld} (Fig. \ref{fig:fig006}) follows the same profile of operations: (1) OS related system calls (recvmsg, futex, poll, wait4), (2) metadata related system calls (stat, fstat, lstat, open, close), (3) user data related system calls (mmap, munmap, read, write). 
Finally, \texttt{Installation} (Fig. \ref{fig:fig007}) distributes the execution activity among: (1) OS related system calls (recvmsg, fcntl, poll), (2) metadata related system calls (stat, open, close), (3) user data related system calls (mmap, munmap, read, write).

%Fig. \ref{fig:fig008} 
%Table~\ref{tab:syscall} summarize the most frequent and the most time-consuming system calls that need to be investigated more deeply.

We summarize the most frequent and the most time-consuming system calls in Table~\ref{tab:syscall}, 
and investigate them further on different platforms in the next section.

%\section{System Calls Analysis}
\section{Cross-Platform Comparison}

\subsection{Memory operations}

Memory operations is the group of system calls that are responsible for interaction with OS's memory subsystem. The frequency and time consumption analysis revealed the dominating importance of mprotect(), brk(), mmap(), munmap(), and madvise() system calls. The method of memory-mapped file I/O is the hottest topic now. As a result, the mmap() and munmap() histograms were selected for detailed analysis. However, it is worth to point out that, fundamentally, the histograms of rest system calls (mprotect(), brk(), and madvise()) look similar.
\begin{figure}[h]
\centering
 
 \includegraphics[width=0.70\columnwidth,keepaspectratio]{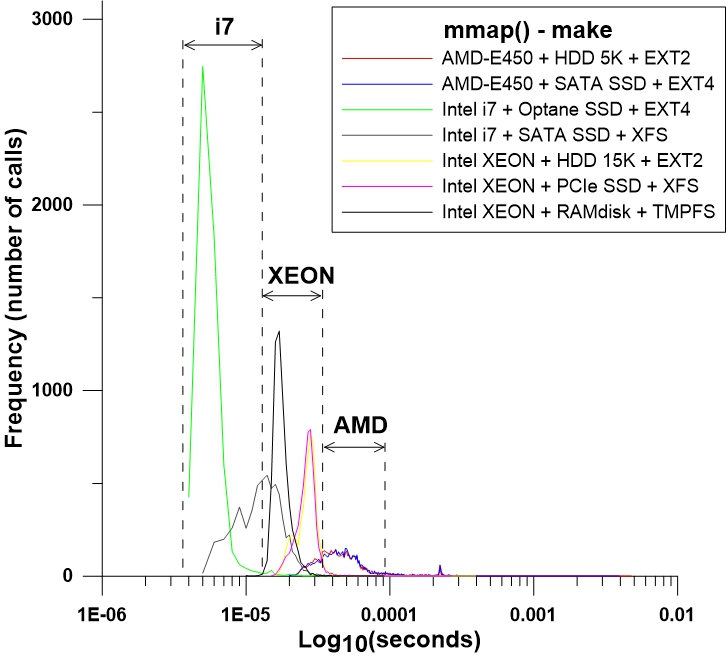}
\caption{Histograms of mmap() system call for the make use-case.}
\label{fig:fig009}
\end{figure}
\begin{figure}[h]
\centering
 
 \includegraphics[width=0.70\columnwidth,keepaspectratio]{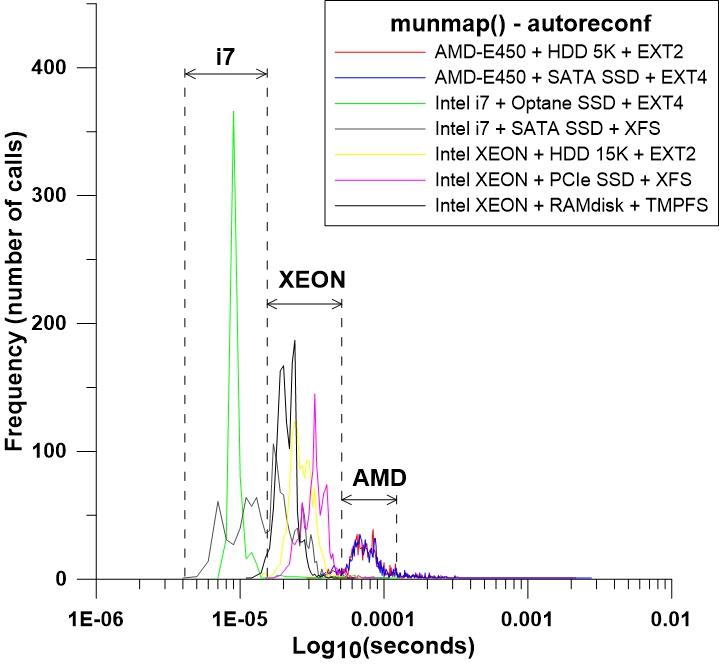}
\caption{Histograms of munmap() system call for the autoreconf use-case.}
\label{fig:fig010}
\end{figure}
\begin{figure}[h]
\centering
 
 \includegraphics[width=0.70\columnwidth,keepaspectratio]{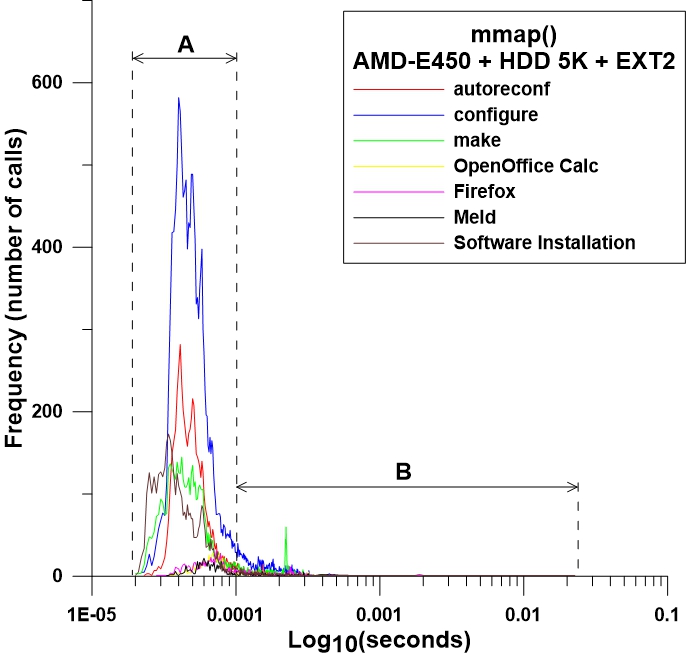}
\caption{Histograms of mmap() system call for AMD + HDD 5K + EXT2.}
\label{fig:fig011}
\end{figure}
\begin{figure}[h]
\centering
 
 \includegraphics[width=0.70\columnwidth,keepaspectratio]{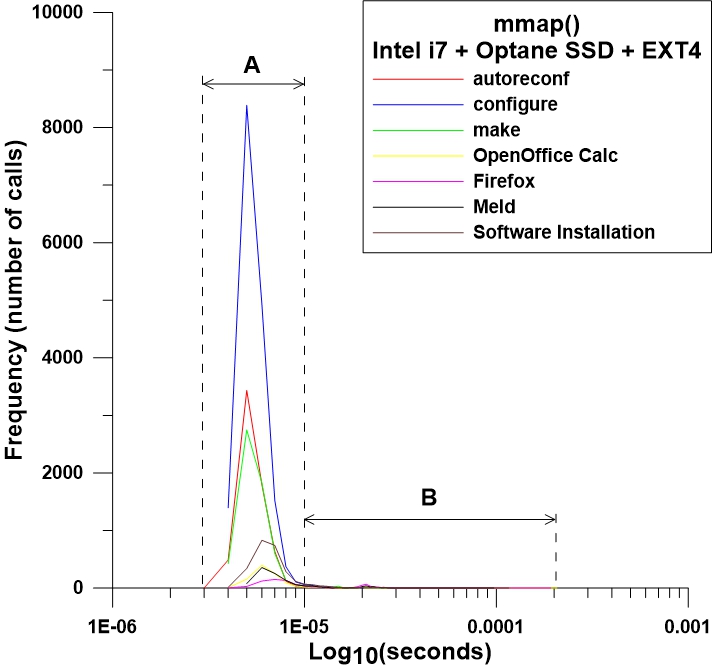}
\caption{Histograms of mmap() system call for i7 + OptaneSSD + EXT4.}
\label{fig:fig012}
\end{figure}
Fig. \ref{fig:fig009} - Fig. \ref{fig:fig010} do not show any valuable difference in histograms for the case of HDD and SSD persistent storages. All such histograms have the same set of peaks. However, HDD-based histograms have the longer tail distribution. It's worth to mention that RAMdisk and OptaneSSD cases change the fine structure of histogram by means of transformation into sharper peak(s) with larger intensity. This peak is moved into the lower latency area. But the important point is that tail distribution is not changed significantly for the RAMdisk and OptaneSSD cases. Sometimes, tail distribution can be longer for OptaneSSD case than for SATA SSD storage device. The really unexpected discovery is the presence of main peak of OptaneSSD in the lower latency area comparing with RAMdisk case. However, it needs to take into account that histograms of OptaneSSD and RAMdisk were built for different platforms (i7 and XEON). Generally speaking, probably, CPU architecture could be more important than persistent memory. Also it's very interesting to compare the SATA SSD's and OptaneSSD's histograms for i7 based platform. It is possible to see that fine structure of OptaneSSD's histogram tends to repeat the fine structure of SATA SSD's histogram. Sometimes, it is possible to see the visible difference in the fine structure of SATA SSD and OptaneSSD. But, very frequently, it is possible to distinguish the same set of peaks for both cases. However, the OptaneSSD's histogram is shifted into the lower latency area, usually. But Fig. \ref{fig:fig010} shows very interesting case when the main peak of OptaneSSD's histogram is surrounded by peaks of SATA SSD's histogram. Fig. \ref{fig:fig011} - Fig. \ref{fig:fig012} show that different use-cases have similar fine structure of histograms. It is possible to see only difference in intensity and length of the tail distribution. Probably, the length of tail distribution is defined by system's background activity. It means that background processes in the system affect the timing of system calls of investigated use-cases. Fig. \ref{fig:fig011} - Fig. \ref{fig:fig012} reveal that even RAMdisk and OptaneSSD have significant length of the tail distribution. It is possible to conclude that faster persistent storage is unable to exclude the tail distribution. Most probably, the histogram's tail distribution is the fundamental factor for von Neumann computing architecture. It's worth to mention that, for example, brk() is the system call is working with memory subsystem by means of allocation/deallocation the memory in the system. From one point of view, it makes sense to expect that histogram of mprotect(), brk(), mmap(), munmap(), and madvise() should have the simple fine structure in the form of one peak. However, the histograms' fine structure of these system calls contains significant amount of peaks. Moreover, changing the CPU architecture is able to change the histogram's fine structure significantly. Oppositely, changing the persistent storage is unable to change the histogram's fine structure. Only RAMdisk and OptaneSSD cases are able to reduce the fine structure to one intensive peak. However, OptaneSSD case is able to show the fine structure with significant amount of details. Probably, the task scheduler affects the histogram's fine structure for the mprotect(), brk(), mmap(), munmap(), and madvise() system calls.

\subsection{Signal processing}

\begin{figure}[h]
\centering
 
 \includegraphics[width=0.70\columnwidth,keepaspectratio]{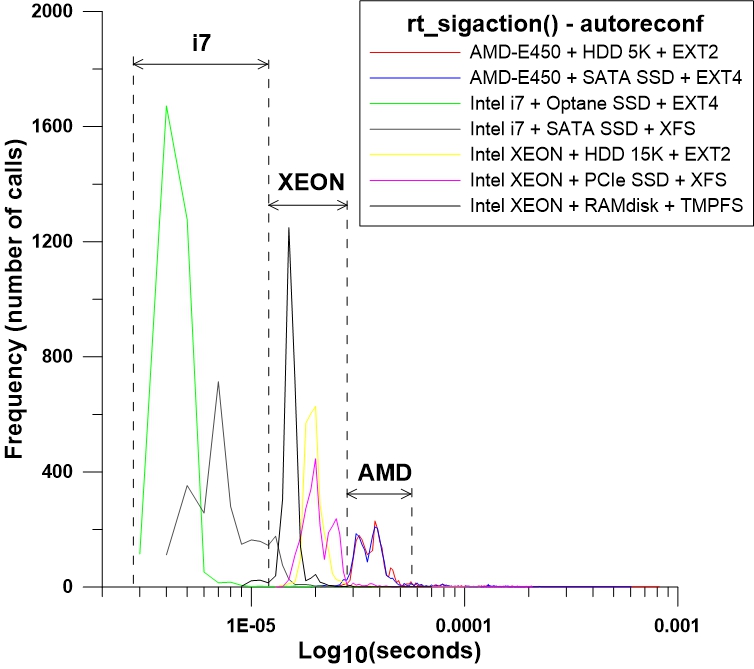}
\caption{Histograms of rt\_sigaction() system call for the autoreconf use-case.}
\label{fig:fig013}
\end{figure}

\begin{figure}[h]
\centering
 
 \includegraphics[width=0.70\columnwidth,keepaspectratio]{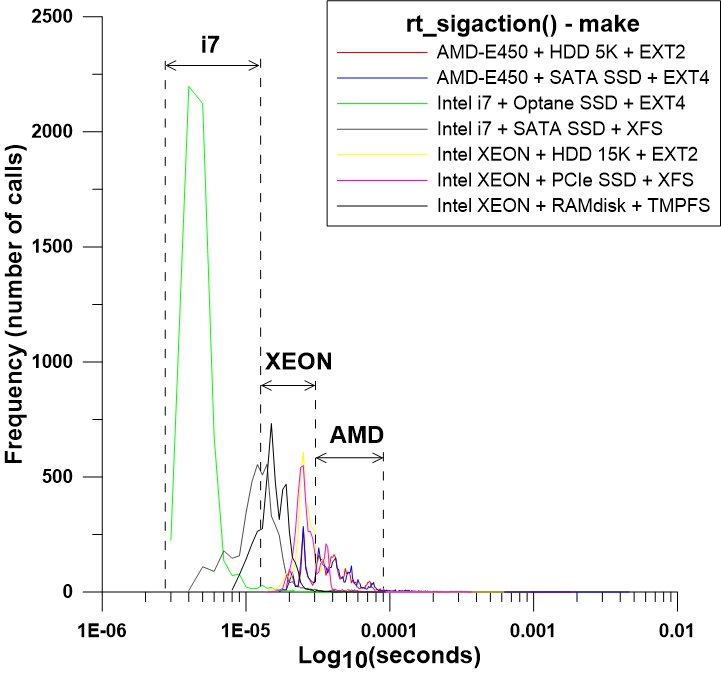}
\caption{Histograms of rt\_sigaction() system call for the make use-case.}
\label{fig:fig014}
\end{figure}

The sigaction() system call is used to change the action taken by a process on receipt of a specific signal. The rt\_sigaction() system call looks better for the case of XEON + RAMdisk and i7 + OptaneSSD (see Fig. \ref{fig:fig013} - Fig. \ref{fig:fig014}).
\begin{figure}[h]
\centering
 
 \includegraphics[width=0.70\columnwidth,keepaspectratio]{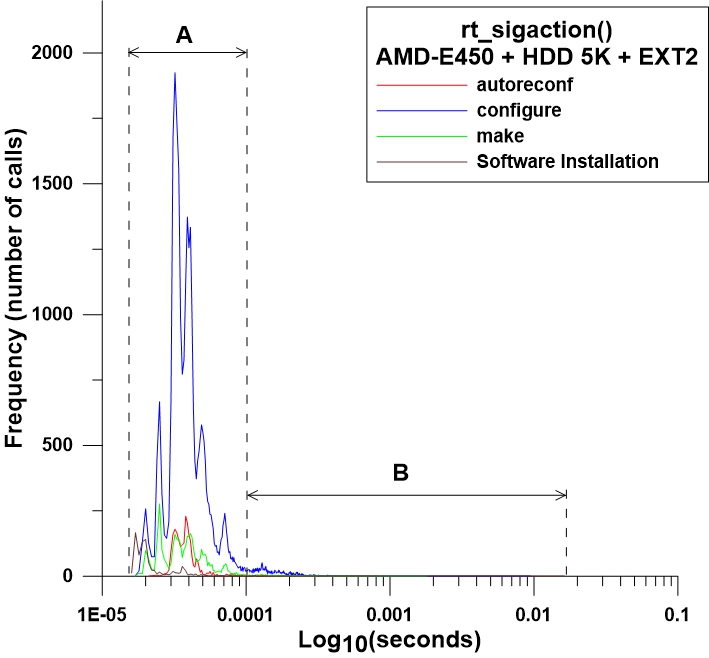}
\caption{Histograms of rt\_sigaction() system call for AMD + HDD 5K + EXT2.}
\label{fig:fig015}
\end{figure}
\begin{figure}[h]
\centering
 
 \includegraphics[width=0.70\columnwidth,keepaspectratio]{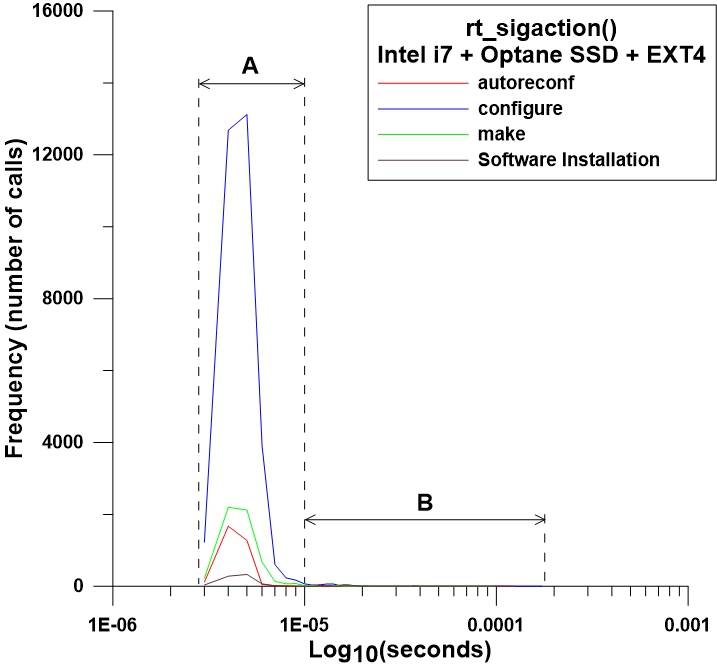}
\caption{Histograms of rt\_sigaction() system call for i7 + OptaneSSD + EXT4.}
\label{fig:fig016}
\end{figure}
However, histogram of i7 + OptaneSSD platform contains the main peak in the lower latencies area comparing with XEON + RAMdisk case. This fact can be considered as a basis for the conclusion that CPU architecture plays more important role than persistent memory in CPU-centric data processing paradigm. However, the type of persistent storage/memory is able to improve slightly the application's performance for the same CPU architecture. But Fig. \ref{fig:fig015} - Fig. \ref{fig:fig016} show that such improvement cannot be significant because even OptaneSSD is unable to shrink the tail distribution significantly. It is possible to imagine the signals like software interrupts. When a signal is sent to a process or thread, a signal handler may be entered, which is similar to the system entering an interrupt handler as the result of receiving an interrupt. Generally speaking, the signals can be treated like synchronization primitive. As a result, the signal-based inter-process communication cannot be improved by a new type of persistent memory because the CPU architecture plays more crucial role in CPU-centric data processing paradigm. 

\subsection{Interprocess communication}

Interprocess communication (IPC) is a set of programming interfaces that allow a programmer to coordinate activities among different program processes that can run concurrently in an operating system. The frequency and time consumption analysis revealed the importance of pipe(), recvmsg(), recvfrom() system calls. The pipe() system call creates a unidirectional data channel that can be used for interprocess communication. Data written to the write end of the pipe is buffered by the kernel until it is read from the read end of the pipe.
\begin{figure}[h]
\centering
 
 \includegraphics[width=0.70\columnwidth,keepaspectratio]{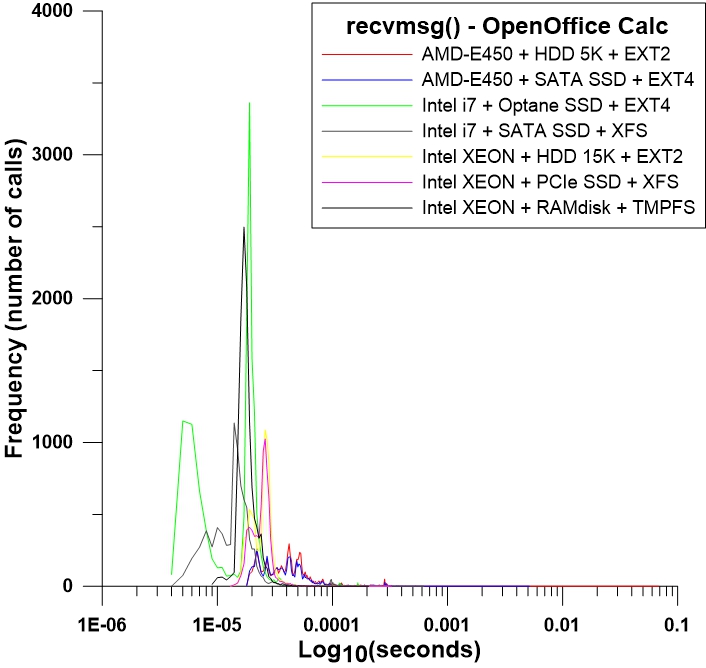}
\caption{Histograms of recvmsg() system call for the OpenOffice Calc use-case.}
\label{fig:fig017}
\end{figure}
\begin{figure}[h]
\centering
 
 \includegraphics[width=0.70\columnwidth,keepaspectratio]{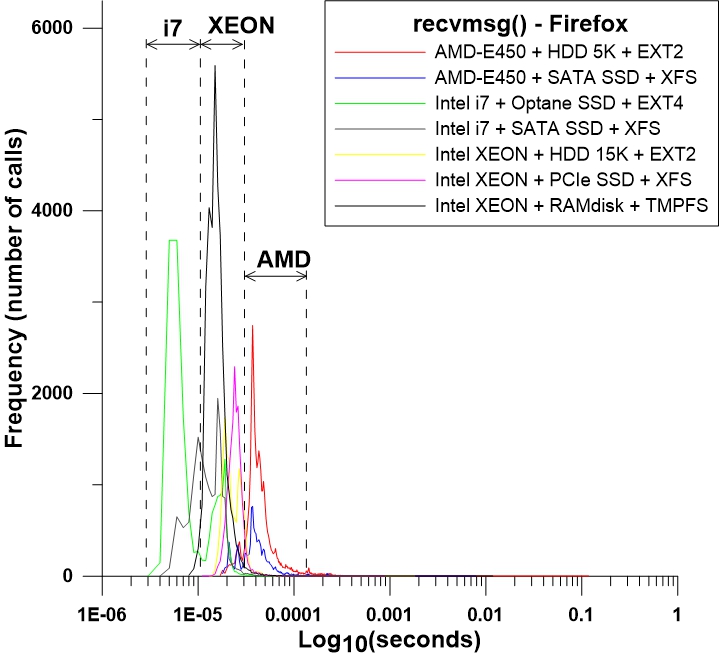}
\caption{Histograms of recvmsg() system call for the Firefox use-case.}
\label{fig:fig018}
\end{figure}
Fig. \ref{fig:fig017} - Fig. \ref{fig:fig018} do not show any visible improvement of recvmsg() system call's execution time for the RAMdisk and OptaneSSD cases. The i7 + SATA SSD easily competes with XEON + RAMdisk and i7 + OptaneSSD. Even XEON + {HDD 15K, PCIe SSD} are able to compete with XEON + RAMdisk for the case of recvmsg() system call. Also OptaneSSD case is unable to shrink the tail distribution significantly (see Fig. \ref{fig:fig017} - Fig. \ref{fig:fig018}). The recvmsg() system call implements the inter-process communications and it represents very important OS's mechanism that affects the applications' performance significantly. But available results do not show any significant improvement for the case of recvmsg() system call. But it is possible to see the same sequence AMD-\textgreater XEON-\textgreater i7 where the histograms for i7 platform are located in the area with lowest latency values. It looks like that CPU architecture plays the most important role in defining the efficiency of recvmsg() system call. The inter-process communication is the cornerstone of improving the applications performance by means of parallel execution of applications' sub-tasks. But persistent memory is unable to improve the performance of inter-process communications for the CPU-centric data processing paradigm.

\subsection{File operations}

The file operations are metadata related system calls. Any user-space application uses the metadata related system calls with significant frequency. Frequently, these type of system calls dominate in the application profile. The frequency and time consumption analysis revealed the dominating importance of dup2(), stat(), fstat(), lstat(), open(), close(), access(), lseek(), fcntl() system calls.
\begin{figure}[h]
\centering
 
 \includegraphics[width=0.70\columnwidth,keepaspectratio]{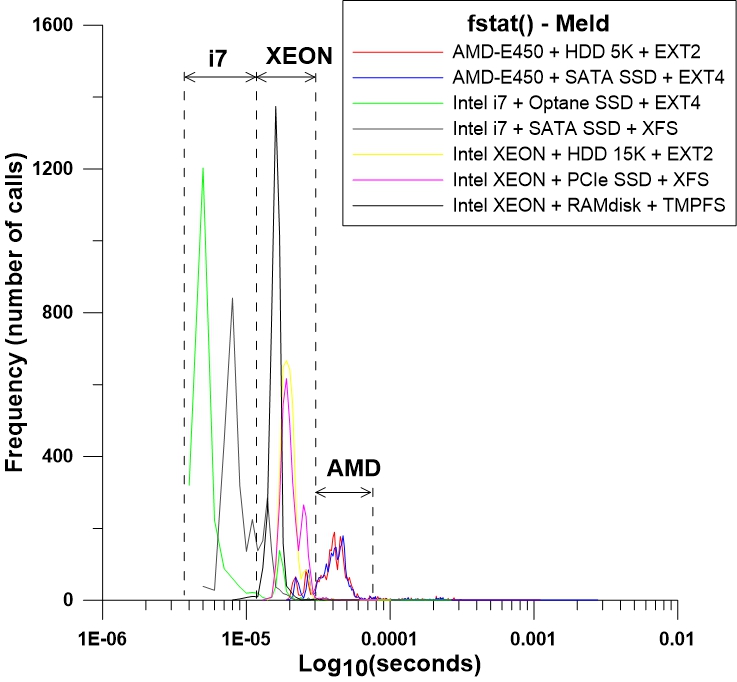}
\caption{Histograms of fstat() system call for the Meld use-case.}
\label{fig:fig019}
\end{figure}
\begin{figure}[h]
\centering
 
 \includegraphics[width=0.70\columnwidth,keepaspectratio]{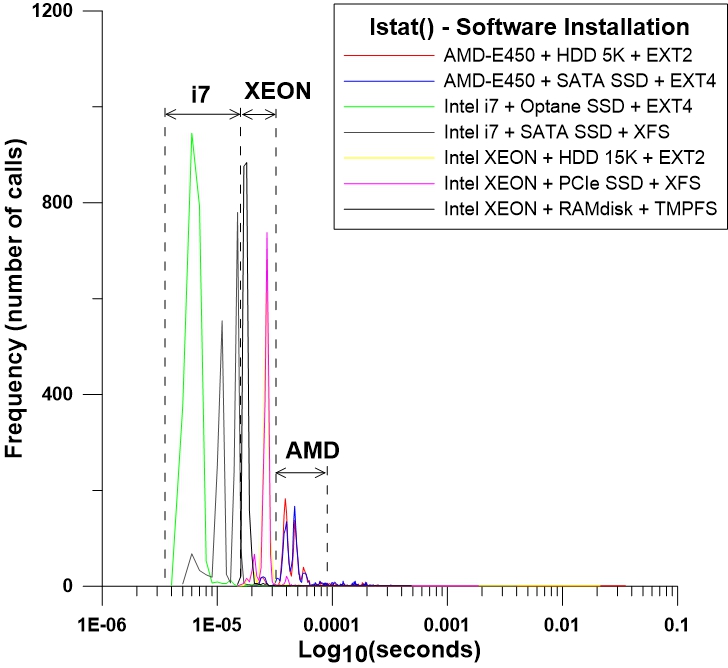}
\caption{Histograms of lstat() system call for the software installation use-case.}
\label{fig:fig020}
\end{figure}
It is possible to see that Fig. \ref{fig:fig019} - Fig. \ref{fig:fig020} do not show any principal difference in histograms for the AMD + HDD 5K and AMD + SATA SSD. The same situation can be concluded for the XEON + HDD 15K and XEON + PCIe SSD (see Fig. \ref{fig:fig019} - Fig. \ref{fig:fig020}). Even if the stat() system calls family is dependent from operations with persistent storage but it is not possible to see that type of persistent storage is able to improve the performance dramatically. The histograms show only one difference – the length of the tail distribution. But it was discovered that HDD 15K has shorter tail distribution comparing with PCIe SSD. Generally speaking, the type of persistent storage is not steady basis for improving the application's performance at whole. The XEON + RAMdisk case looks better for the software installation use-case (see Fig. \ref{fig:fig020}). But another use-cases (Meld and Firefox, for example) do not show any significant advantages for the RAMdisk case (see Fig. \ref{fig:fig019}). The i7 + OptaneSSD case looks better than i7 + SATA SSD but it is possible to state that SATA SSD is able to compete with OptaneSSD. Finally, generally speaking, the persistent memory is unable to improve the total application's performance alone. Fig. \ref{fig:fig019} - Fig. \ref{fig:fig020} clearly show that CPU architecture is more influential factor than type of persistent memory.

\subsection{Locking operations}

The locking primitives are used in multi-threaded applications very frequently because of necessity to synchronize the access to shared data. The frequency and time consumption analysis revealed the dominating importance of futex(), poll(), and wait4() system calls. The futex() system call has very special histograms. First of all, all histograms have very long tail distribution. And different persistent storage types are unable to shrink or to exclude the tail distribution. Moreover, the tail distribution is practically unchanged for any CPU architecture or persistent storage (see Fig. \ref{fig:fig023} - Fig. \ref{fig:fig024}).
\begin{figure}[h]
\centering
 
 \includegraphics[width=0.70\columnwidth,keepaspectratio]{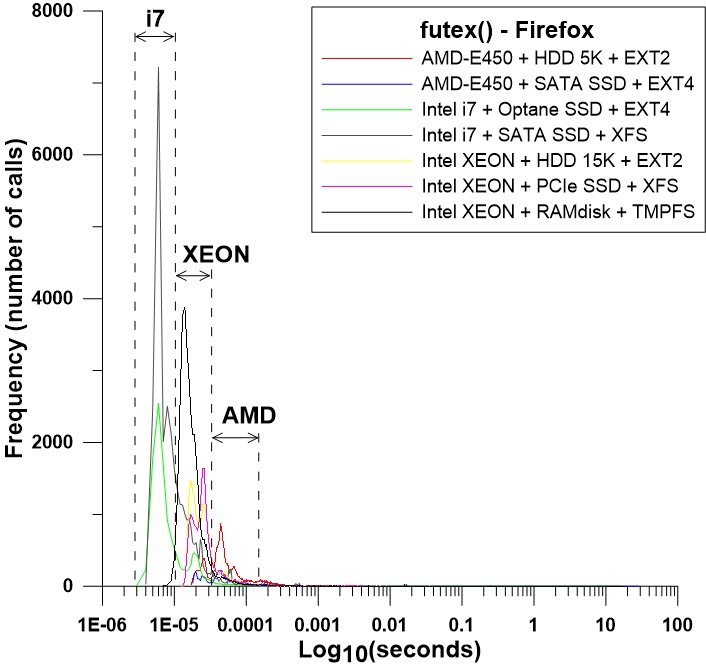}
\caption{Histograms of futex() system call for the Firefox use-case.}
\label{fig:fig021}
\end{figure}
\begin{figure}[h]
\centering
 
 \includegraphics[width=0.70\columnwidth,keepaspectratio]{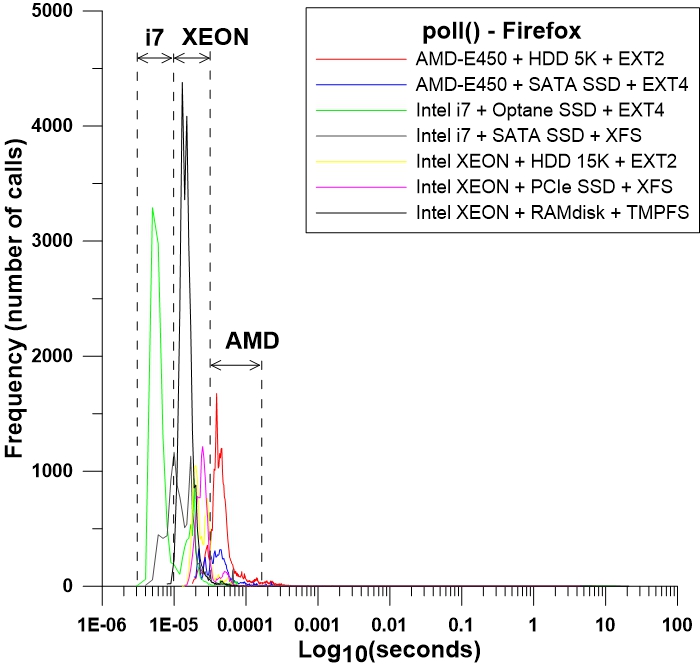}
\caption{Histograms of poll() system call for the Firefox use-case.}
\label{fig:fig022}
\end{figure}
\begin{figure}[h]
\centering
 
 \includegraphics[width=0.70\columnwidth,keepaspectratio]{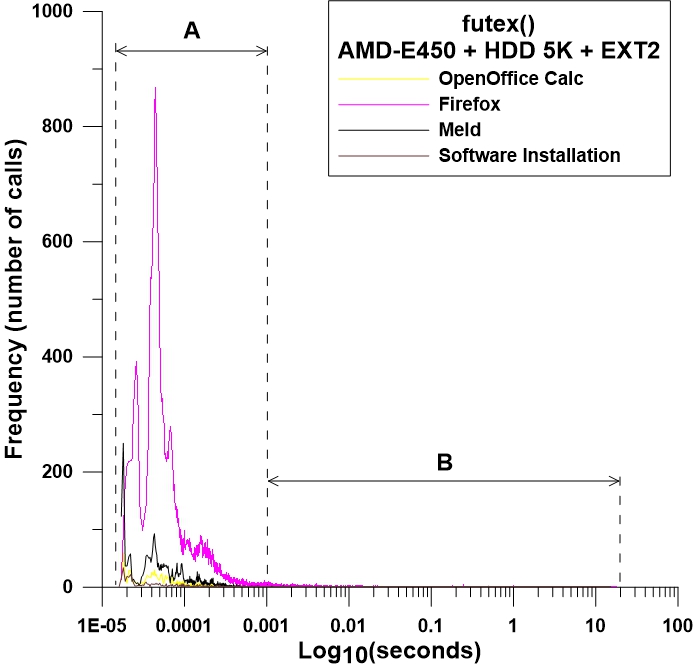}
\caption{Histograms of futex() system call for AMD + HDD 5K + EXT2.}
\label{fig:fig023}
\end{figure}
\begin{figure}[h]
\centering
 
 \includegraphics[width=0.70\columnwidth,keepaspectratio]{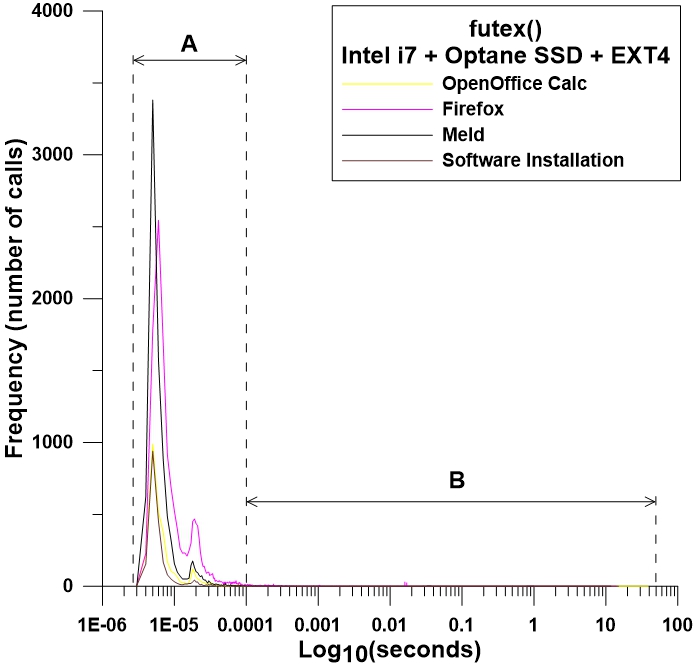}
\caption{Histograms of futex() system call for i7 + OptaneSSD + EXT4.}
\label{fig:fig024}
\end{figure}
Even RAMdisk and OptaneSSD have really long tail distribution that affects the total execution time of the use-cases. Generally speaking, main peaks of all histograms are grouped in a narrow area (see Fig. \ref{fig:fig021}). There is only one difference – redistribution of the peaks' intensity among different latency values. However, it is easy to see the same sequence AMD-\textgreater XEON-\textgreater i7 where the histograms for i7 platform are located in the area with lowest latency values (see Fig. \ref{fig:fig021}). The futex() system call is one that significantly affects the total execution time. Fig. \ref{fig:fig022} show that all investigated use-cases are unable to improve average execution time or to shrink the tail distribution of poll() system call. The poll() system call plays the role of synchronization primitive is associated with I/O operations. And it is easy to see that even RAMdisk and OptaneSSD are unable to change anything. Generally speaking, the synchronization primitives degrades the whole performance significantly but the von Neumann paradigm doesn't provide any hope to resolve the drawback. There are some fundamental issues in CPU-centric paradigm and modern OS architectures. Everybody believes that the next generation of NVM/SCM memory is able to improve the performance/latency of read/write operations significantly. It means that average execution time of I/O operations should be reduced. As a result, an application's threads should spend much lesser in synchronization primitives that wait the ending of I/O operations. However, as RAMdisk as OptaneSSD cases are unable to show any significant improvement of synchronization primitives' as average as total execution time. Most probable, the key drawbacks take place in CPU architecture, task scheduler implementation and the whole interaction of OS subsystems during the processes/threads management. Also, the memory management subsystem is able to play very important role that is able to affect the performance of I/O operations and synchronization primitives. The available results provides the basis for the conclusion that the changing as persistent storage/memory as CPU architecture is unable to decrease the average execution time of futex(), poll(), and wait4() system calls. Most probably, the von Neumann paradigm has fundamental drawback(s) that cannot be resolved by faster CPU or faster persistent memory.

\subsection{User data operations}

The read(), write() system calls are cornerstones for operation with user data. The read() system call is the very critical function that directly defines the performance of operations with user data. Another important point that read() is executed synchronously by the kernel, usually. Naïve point of view provides expectation that RAMdisk or OptaneSSD have to improve the performance of read() system call dramatically.
\begin{figure}[h]
\centering
 
 \includegraphics[width=0.70\columnwidth,keepaspectratio]{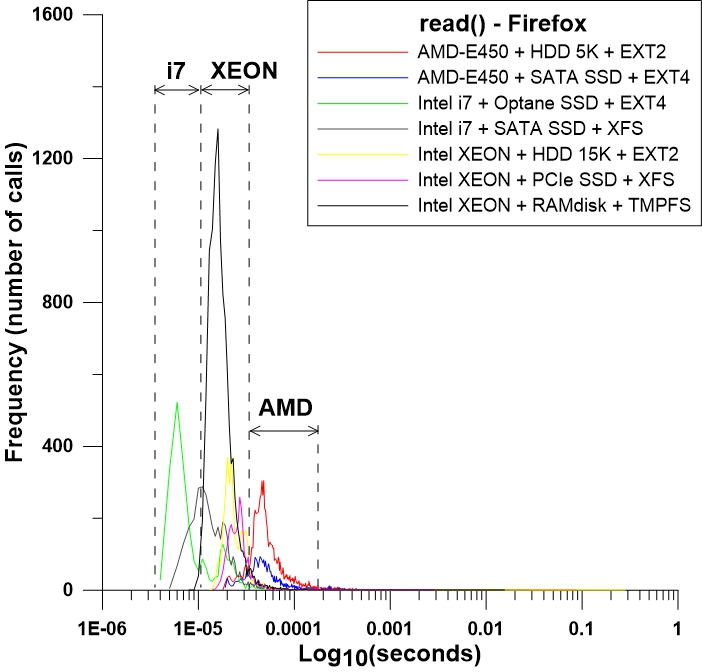}
\caption{Histograms of read() system call for the Firefox use-case.}
\label{fig:fig025}
\end{figure}
\begin{figure}[h]
\centering
 
 \includegraphics[width=0.70\columnwidth,keepaspectratio]{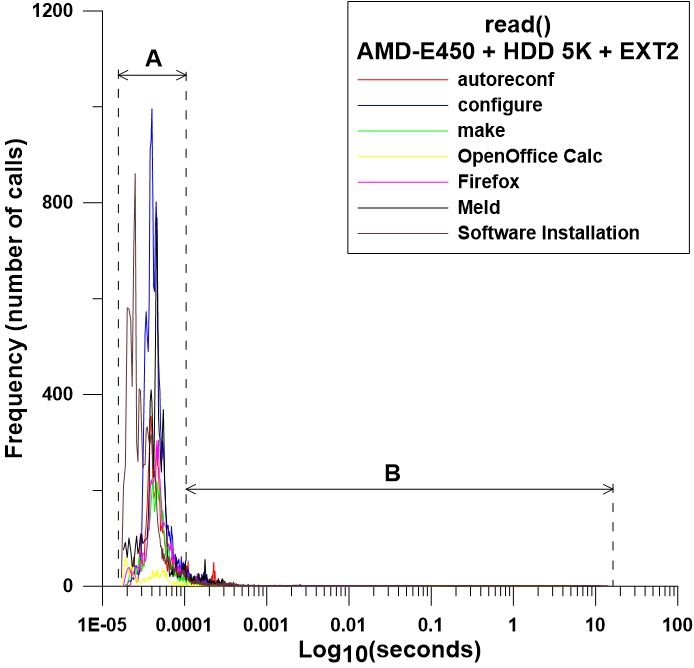}
\caption{Histograms of read() system call for AMD + HDD 5K + EXT2.}
\label{fig:fig026}
\end{figure}
\begin{figure}[h]
\centering
 
 \includegraphics[width=0.70\columnwidth,keepaspectratio]{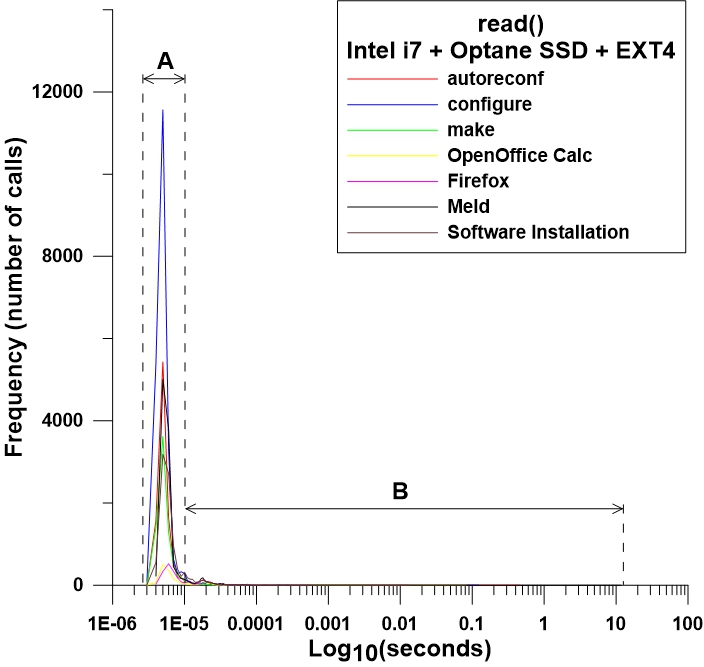}
\caption{Histograms of read() system call for i7 + OptaneSSD + EXT4.}
\label{fig:fig027}
\end{figure}
\begin{figure}[h]
\centering
 
 \includegraphics[width=0.70\columnwidth,keepaspectratio]{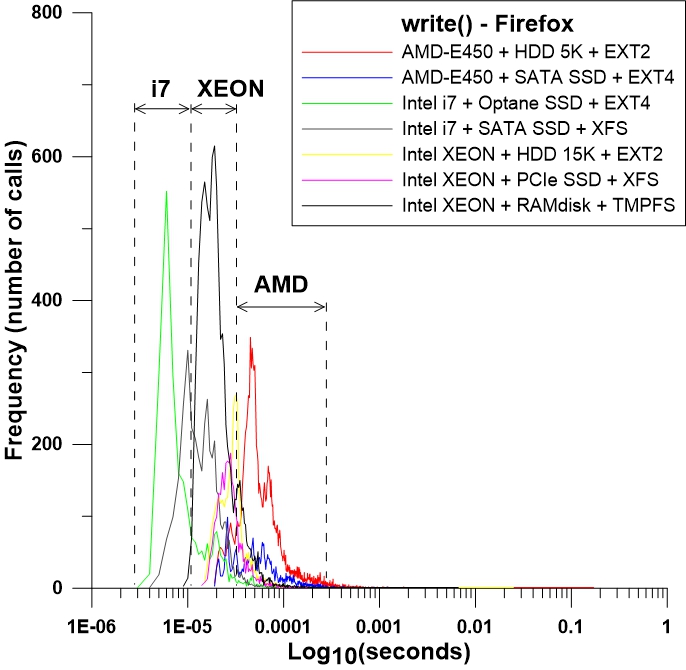}
\caption{Histograms of write() system call for the Firefox use-case.}
\label{fig:fig028}
\end{figure}
\begin{figure}[h]
\centering
 
 \includegraphics[width=0.70\columnwidth,keepaspectratio]{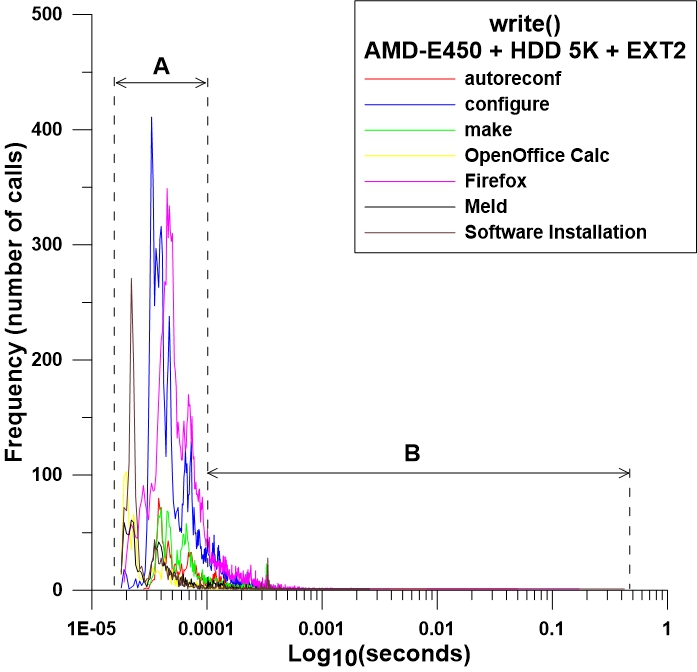}
\caption{Histograms of write() system call for AMD + HDD 5K + EXT2.}
\label{fig:fig029}
\end{figure}
\begin{figure}[h]
\centering
 
 \includegraphics[width=0.70\columnwidth,keepaspectratio]{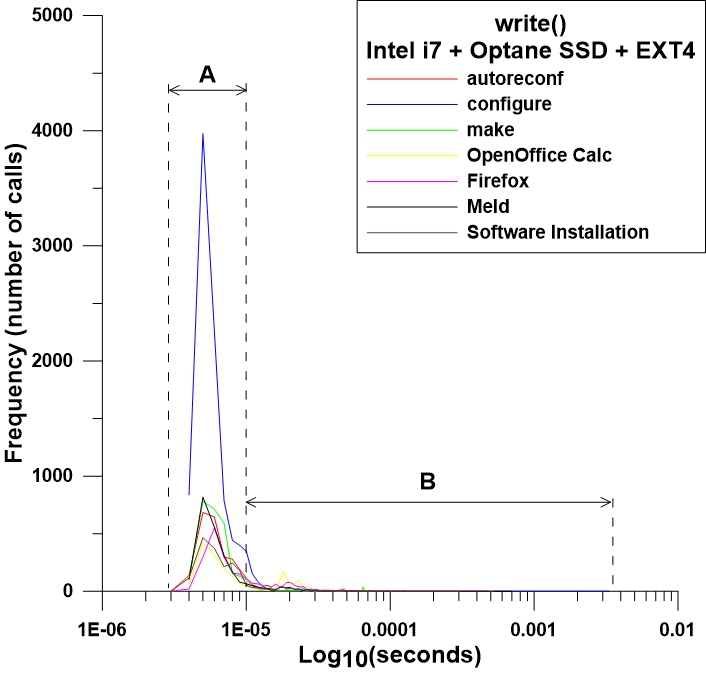}
\caption{Histograms of write() system call for i7 + OptaneSSD + EXT4.}
\label{fig:fig030}
\end{figure}
However, Fig. \ref{fig:fig025} do not show any significant improvements for the read() system call by means of varying the persistent storage type. Finally, all histograms provide the basis for the paradoxical conclusion that CPU architecture is the more influential factor for read() system call. The faster persistent memory is able to shrink the tail distribution slightly but it is unable to change the nature of histogram at whole (see Fig. \ref{fig:fig026} - Fig. \ref{fig:fig027}). Generally speaking, the read() system call also reveals the fundamental influence of CPU architecture in the CPU-centric data processing paradigm instead of expected influence of persistent storage type. Fig. \ref{fig:fig028} does not show any significant improvement comparing HDD and SSD cases for the case of write() system call. It is possible to see only difference in the tail distribution for these cases. The SSD based case is able to shrink the tail distribution slightly. The RAMdisk and OptaneSSD cases are able to simplify the fine structure of the histogram by means of increasing the intensity of the main peak. Also, usually, the main peak is located into lower latencies area comparing with HDD or SSD cases. However, even RAMdisk or OptaneSSD are unable to shrink the tail distribution significantly. The most important point that the write() system call has the histograms of i7 + OptaneSSD are located in lower latencies area comparing with XEON + RAMdisk case.

\section{Discussion}

\textbf{User-data operations}. The read() system call is the very critical function that directly defines the performance of operations with user data. Another important point that read() is executed synchronously by the kernel, usually. Naive point of view provides expectation that RAMdisk or OptaneSSD have to improve the performance of read() system call dramatically. However, histograms do not show any significant improvements for the read() system call by means of varying the persistent storage type. Finally, all histograms provide the basis for the paradoxical conclusion that CPU architecture is the more influential factor for the read() system call.

\textbf{Metadata operations}. It is possible to see that type of persistent storage is unable to improve the performance of metadata-related system calls dramatically (for example, stat() system call) even if the metadata-related system calls are dependent from operations with persistent storage. The histograms show only one difference – the length of the tail distribution. But it was discovered that HDD 15K could have shorter tail distribution comparing with PCIe SSD. It is possible to see that the tail distribution has very interesting peculiarities for some cases. Such peculiarities have fundamental reasons that could be created by OS background activity or system call nature. The really important point that the tail distribution cannot be shrunk significantly for the case of CPU-centric data processing architecture. The background activity of POSIX OS is able to eliminate completely the advantages of fast persistent memory. Finally, the RAMdisk and OptaneSSD cases reveal the influence of the OS background activity. Most probably, the CPU-centric data processing is responsible for the impossibility of new type of NVM/SCM memory to improve the system performance at whole for any use-case.

\textbf{Synchronization primitives}. The futex() system call has very special histograms. First of all, all histograms have very long tail distribution. And different persistent storage types are unable to shrink or to exclude the tail distribution. Moreover, the tail distribution is practically unchanged for any CPU architecture or persistent storage. Even RAMdisk and OptaneSSD have really long tail distribution that affects the total execution time of the use-cases. The futex() system call is one that significantly affects the total execution time. The poll() system call plays the role of synchronization primitive is associated with I/O operations. Everybody believes that the next generation of NVM/SCM memory is able to improve the performance/latency of read/write operations significantly. It means that average execution time of I/O operations should be reduced. As a result, an application’s threads should spend much lesser time in synchronization primitives that wait the ending of I/O operations. However, as RAMdisk as OptaneSSD cases are unable to show any significant improvement of synchronization primitives' as average as total execution time. Most probable, the key drawbacks take place in CPU architecture, task scheduler implementation and the whole interaction of OS subsystems during the processes/threads management. Also, the memory management subsystem is able to play very important role that is able to affect the performance of I/O operations and synchronization primitives. Generally speaking, the combination of CPU-centric data processing with POSIX-based OS architecture is affected by CPU architecture more significantly than by persistent memory.

\textbf{Inter-process communication}. The inter-process communication is the cornerstone of improving the applications performance by means of parallel execution of applications' sub-tasks. But persistent memory is unable to improve the performance of inter-process communications for the CPU-centric data processing paradigm. It is possible to summarize that CPU architecture defines the position of histograms for different platforms because even XEON + RAMdisk platform is unable to compete with i7 + {OptaneSSD; SATA SSD} cases.

\textbf{CPU-centric architecture}. The really unexpected discovery is the presence of main peak of OptaneSSD in the lower latency area comparing with RAMdisk case. Also, the histograms reveal that even RAMdisk and OptaneSSD have significant length of the tail distribution. It is possible to conclude that faster persistent storage is unable to exclude the tail distribution. Most probably, the histogram's tail distribution is the fundamental factor for von Neumann computing architecture. Moreover, changing the CPU architecture is able to change the histogram's fine structure significantly. Oppositely, changing the persistent storage is unable to change the histogram's fine structure. Generally speaking, the type of persistent memory or storage is not steady basis for improving the performance of application. Even the tail distribution can be shrunk more efficiently by means of changing the CPU architecture but not the persistent memory type. However, the type of persistent storage/memory is able to improve slightly the application's performance for the same CPU architecture.

\section{Conclusion}

The next generation of NVM/SCM memory would be able to open the new horizons for future computer technologies. However, currently, the NVM/SCM memory represents the big challenge but not the hope to resolve the computer science's problems. We've built histograms for the most frequent and time-consuming system calls with the goal to understand the nature of distribution for different platforms. It was discovered unexpected and stable dependence of histograms from the CPU architecture. However, the type of persistent storage doesn't play important role in the discovered dependence. Different use-cases have similar fine structure of histograms. It is possible to see only difference in intensity and length of the tail distribution. Probably, the length of tail distribution is defined by system's background activity. It means that background processes in the system affect the timing of system calls of investigated use-cases. The analysis of histograms showed that faster persistent storage is unable to exclude the tail distribution. Most probably, the histogram's tail distribution is the fundamental factor for von Neumann computing architecture. It was discovered that changing the CPU architecture is able to change the histogram's fine structure significantly. Oppositely, changing the persistent storage is unable to change the histogram's fine structure. Only RAMdisk and OptaneSSD cases are able to reduce the fine structure to one intensive peak. However, OptaneSSD case is able to show the fine structure with significant amount of details. Probably, the task scheduler affects the histogram's fine structure. This fact can be considered as a basis for the conclusion that CPU architecture plays more important role than persistent memory in CPU-centric data processing paradigm. However, the type of persistent storage/memory is able to improve slightly the application's performance for the same CPU architecture. The inter-process communication is the cornerstone of improving the applications performance by means of parallel execution of applications' sub-tasks. But the signal-based inter-process communication cannot be improved by a new type of persistent memory because the CPU architecture plays more crucial role in CPU-centric data processing paradigm. The synchronization primitives degrades the whole performance significantly but the von Neumann paradigm doesn't provide any hope to resolve the drawback. Most probable, the key drawbacks take place in CPU architecture, task scheduler implementation and the whole interaction of OS subsystems during the processes/threads management. Also, the memory management subsystem is able to play very important role that is able to affect the performance of I/O operations and synchronization primitives.


%%%%%%%%%%%%%%%%%%%%%%%%%%%%%%%%%%%%%%%%%%%%%%%%%%%%%%%%%%%%%%%%%%%%%%%%%%%%%%%%

%%%%%%%%%%%%%%%%%%%%%%%%%%%%%%%%%%%%%%%%%%%%%%%%%%%%%%%%%%%%%%%%%%%%%%%%%%%%%%%%

%%%%%%%%%%%%%%%%%%%%%%%%%%%%%%%%%%%%%%%%%%%%%%%%%%%%%%%%%%%%%%%%%%%%%%%%%%%%%%%%

%%%%%%%%%%%%%%%%%%%%%%%%%%%%%%%%%%%%%%%%%%%%%%%%%%%%%%%%%%%%%%%%%%%%%%%%%%%%%%%%

\end{document}